\documentclass[aps,prl,twocolumn,superscriptaddress]{revtex4-1}
\usepackage{graphicx}
\usepackage{epstopdf}
\usepackage{color}
\usepackage{bm}
\usepackage{amsmath, amssymb, physics}
\usepackage{changes}
\usepackage[colorlinks=true,linkcolor=blue,urlcolor=blue,citecolor=blue]{hyperref}
\usepackage[mathlines]{lineno}% Enable numbering of text and display math
\usepackage{cancel}
\usepackage{ulem}

\definecolor{pink}{rgb}{1,1,0} % color values Red, Green, Blue
\definecolor{red}{rgb}{1,0,0}
\definecolor{yellow}{rgb}{1,1,0}
\definecolor{orange}{rgb}{1,0.5,0}
\definecolor{green}{rgb}{0,1,0}
\definecolor{blue}{rgb}{0,0,1}
\definecolor{white}{rgb}{1,1,1}
\definecolor{purple}{rgb}{0.5,0,0.5}

\begin{document}
\title{Excitonic Charge Density Waves in Moire Ladders}
\author{Paula Mellado}
 \affiliation{Facultad de Ingeniería y Ciencias,\\ Universidad Adolfo Ibáñez, Santiago, Chile.}
\author{Francisco Muñoz }%
 \author{Javiera Cabezas-Escares}
\affiliation{
Departamento de Física, Facultad de Ciencias,\\ Universidad de Chile and CEDENNA\\
Santiago, Chile
}%
\begin{abstract} 
An incommensurate charge density wave (CDW) is a periodic modulation of charge that breaks translational symmetry incongruently with the underlying lattice. Its low-energy excitations, the phason, are collective, gapless phase fluctuations. We study a half-filled, four-band ladder model where a shift \(\delta = p/q\) between the legs leads to a supercell of \(q\) composite cells. The moiré potential narrows minibands near the Fermi level, resulting in additional peaks in the density of states, whose separation is controlled by \(\delta\). The inclusion of short-range Coulomb interactions leads to an excitonic incommensurate CDW state. We identify the oscillations in its amplitude with a gapped Higgs collective mode and a lowest-energy Goldstone mode, realized by long-lived neutral phasons whose propagation velocity is governed by the shift \(\delta\) and the inter-leg tunneling amplitude. Our results show that even the slightest interlayer mismatches can strongly modify both charge-ordering patterns and low-energy bosonic excitations in layered materials, and suggest that the enigmatic CDW phase in the quasi-one-dimensional compound \(\rm HfTe_3 \) is excitonic in nature.
\end{abstract}
\maketitle
\emph{Introduction.} 
In real materials, layer mismatch frequently stems from subtle differences between intra- and interlayer bonding \cite{abbas2020recent}. Such incongruity can be generated by relative layer twisting \cite{chen2019tunable} or by pressure- and strain-induced lattice distortions \cite{gao2020band,escudero2024designing, rakib2022moire,gao2022symmetry}. A common effect in low-dimensional systems \cite{mellado2024quantum} is susceptibility to charge density waves \cite{mcmillan1975landau,takada1985damping,gruner1988dynamics,tucker1988theory}, a periodic modulation of the conduction-electron charge density in a crystal \cite{gor2012charge}. This phenomenon occurs primarily in low-dimensional conductors, such as layered crystals \cite{rossnagel2011origin,zhu2015classification,imada1986competition}.  In the past few decades, there have been significant efforts to understand the driving mechanisms of CDW-hosting systems, such as excitonic insulators, quasi-two-dimensional transition metal dichalcogenides, and rare-earth tritellurides ($\rm RTe_3s$), and quasi-one-dimensional (quasi-1D) systems \cite{banerjee2013charge,melikyan2005model,lee1978dynamics,efetov1977charge,littlewood1982amplitude,wegner2020evidence,guo2024real}. 

The emergence of CDWs is frequently ascribed to an electronic instability induced by electron-phonon coupling—the Peierls mechanism \cite{urban2007scaling,hirsch1983effect,boriack1977dynamic}. Alternatively, in excitonic CDW systems, Coulomb interactions play a key role, as they dominate over kinetic energy \cite{gao2024observation}. It has been shown that the observed charge ordering phase transitions are far from being analogs of the Peierls mechanism \cite{johannes2008fermi} because electronic instabilities are easily destroyed by even the slightest deviations from the perfect nesting conditions of the Peierls scenario \cite{hofmann2019strong}. This is the case in the CDW materials $\rm NbSe_2$, \cite{lian2018unveiling}, $\rm TaSe_2$, \cite{ryu2018persistent}, and $\rm CeTe_3$ \cite{ralevic2016charge,johannes2008fermi}.

In a CDW state, the charge density follows a standing wave pattern, with modulation characterized by the complex order parameter $\rho(x)=|\rho_0|e^{i\theta(x)}$ \cite{gor2012charge}, which comprises the amplitude $\rho_0$ and the phase $\theta$, characterizing the spatial translation of the modulation relative to the underlying crystal lattice. Upon the formation of a CDW, a gap opens in the spectrum, and fluctuations of $|\rho_0|$ and  $\theta$ \cite{tucker1988theory} give rise to collective excitations known as amplitudons and phasons \cite{gruner1988dynamics}. The amplitudon is a gapped excitation \cite{kwon2024dual}, while the phason is gapped if pinned or gapless, which often occurs with the formation of a CDW state that is incongruent with the underlying crystal lattice. The incommensurate nature of the CDW state implies that its wavelength is not an integer multiple of the lattice constant. In this case, the phason becomes the Goldstone excitation associated with the spontaneous breaking of translational symmetry \cite{fisher1985sliding,brazovskii2004pinning}.  
While the overall CDW carries a charge current, in incommensurate CDWs the phason is often charge-neutral \cite{birkbeck2024measuring,ochoa2019moire}. Digging into the origin of this neutrality in layered models \cite{miao2019formation} is crucial for successfully capturing massless phasons in higher-dimensional materials and for understanding collective charge dynamics, which could provide insights into dissipationless transport and its coexistence or competition with superconductivity \cite{wilson1985charge,PhysRevLett.42.1423,shapiro2007observation,fukuyama2020theory}.

In this article, we study a toy model of a heterostructure consisting of a half-filled, four-band ladder with a relative shift between the legs \(\delta=p/q\), which creates a moiré supercell of \(q\) composite cells and a moiré potential that compresses layer bands into flat minibands near the Fermi level. Including Coulomb interactions within a mean-field approach, we find an excitonic incommensurate CDW phase characterized by a rung-odd charge-density modulation, leading to long-lived neutral phason excitations. We show that the moiré phasons of our model are Goldstone modes of the effective interacting theory in which the CDW slips, with a speed controlled by \(\delta\) and the inter-leg tunneling amplitude. 
\begin{figure*}
\centering
\includegraphics[width=\linewidth]{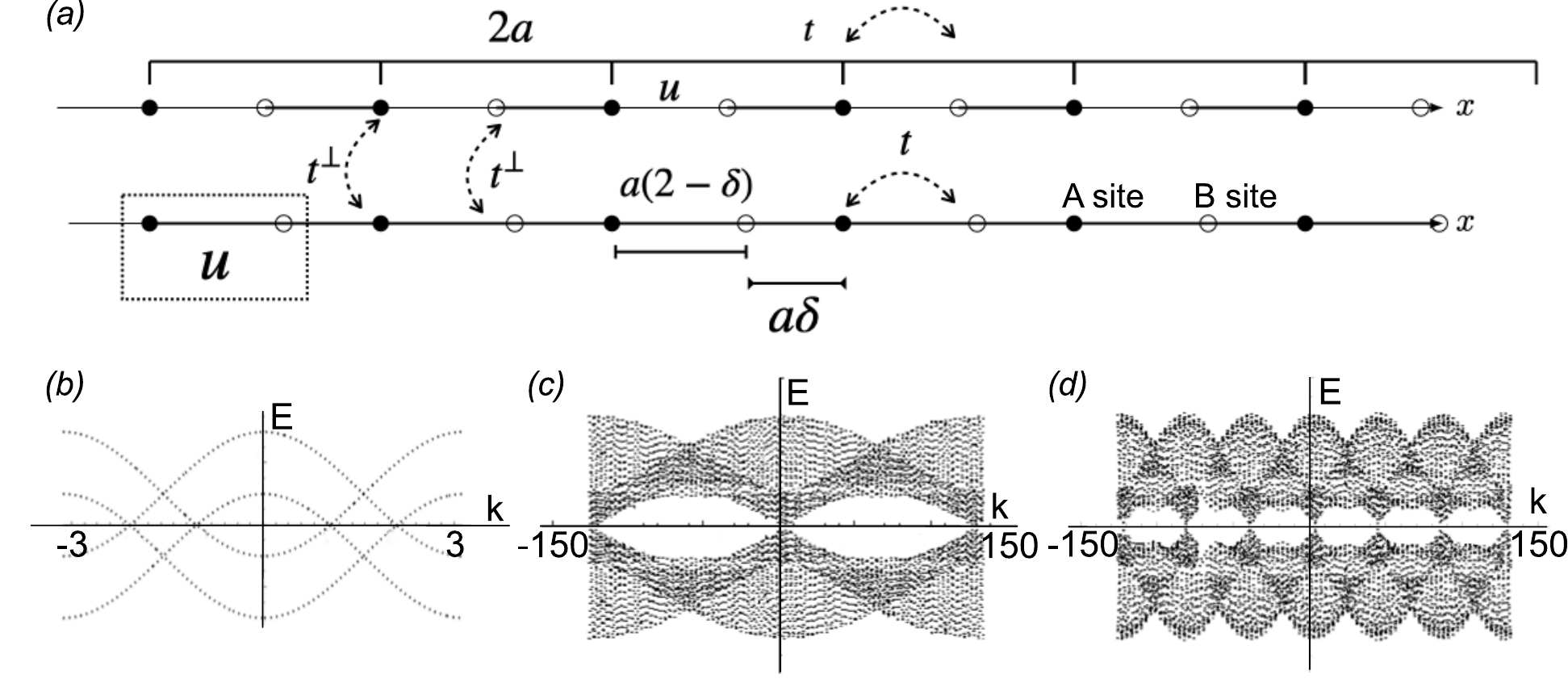}
\caption{\label{f1} (a) Schematic illustration of the effective moiré ladder model. The moiré modulation does not arise from a geometric lattice mismatch but from a long-wavelength modulation of the inter-leg tunneling, representing an emergent symmetry-broken state. (b-d) Band spectra of the kinetic Hamiltonian $H(k)$ in the EBZ at (b) $\delta=1$, (c) $\frac{19}{20}$ and (d) $\frac{17}{20}$. The large momentum range reflects an extended Brillouin zone representation in which all moiré replicas are unfolded.} 
\end{figure*}

\emph{Model}
The ladder of Fig.\ref{f1}(a) consists of two coupled legs, each with two sites (A and B) per unit cell. The composite ladder unit cell contains a total of four sites 
$\{A_1,B_1,A_2,B_2\}$ (subscripts $1,2$ label legs). Throughout, we choose \textit{a}, half the legs lattice constant, as the length unit, so the composite cell has a length equal to $2$, and the original reciprocal period is \(
G_0=\pi\). Leg-1 is uniform, while leg-2 is dimerized by a shift \(\delta \) per cell. This becomes the moiré parameter of the ladder, \(\delta=p/q\in(0,1)\), where p and q are coprime integers. The relative mismatch between legs yields a moiré supercell of \(q\) composite cells with \(4q\) sites in total, a moiré reciprocal vector in the reduced Brillouin zone (RBZ) \(b_s=\frac{G_0}{q}=\frac{\pi}{q}\), and a real-space supercell length \(L_s=2\,q\). The intra-leg nearest neighbor hopping, $t$, sets the bandwidth \(W\sim 4t\). Inter-leg hopping (rung tunneling) occurs between the same sublattice, and its average is \(t\). Schrieffer–Wolff perturbation theory \cite{schrieffer1966relation} gives rise to a first-harmonic rung tunneling of amplitude \(t_1=(1-\delta)t\) modulated by the envelope $t_\perp(n)=t+t_1\cos\left(\frac{2\pi n}{q}\right)$.

In the moiré ladder studied here, both legs have identical microscopic lattice periodicity, so no geometric moiré pattern arises directly at the level of the underlying lattice. Rather, the moiré superstructure is introduced phenomenologically through the long-wavelength modulation of the inter-leg (rung) tunneling amplitude. This modulation can be viewed as an emergent outcome of spontaneously broken translational symmetry, which arises, as our DFPT calculations show below, from a soft phonon mode that produces relative phase shifts or displacements between adjacent chains. In this sense, the moiré ladder serves as an effective low-energy description of a symmetry-broken phase where a long-wavelength modulation has already formed. 

Throughout, we choose \(t\) as our energy scale. 
\begin{figure}
\centering
\includegraphics[width=\linewidth]{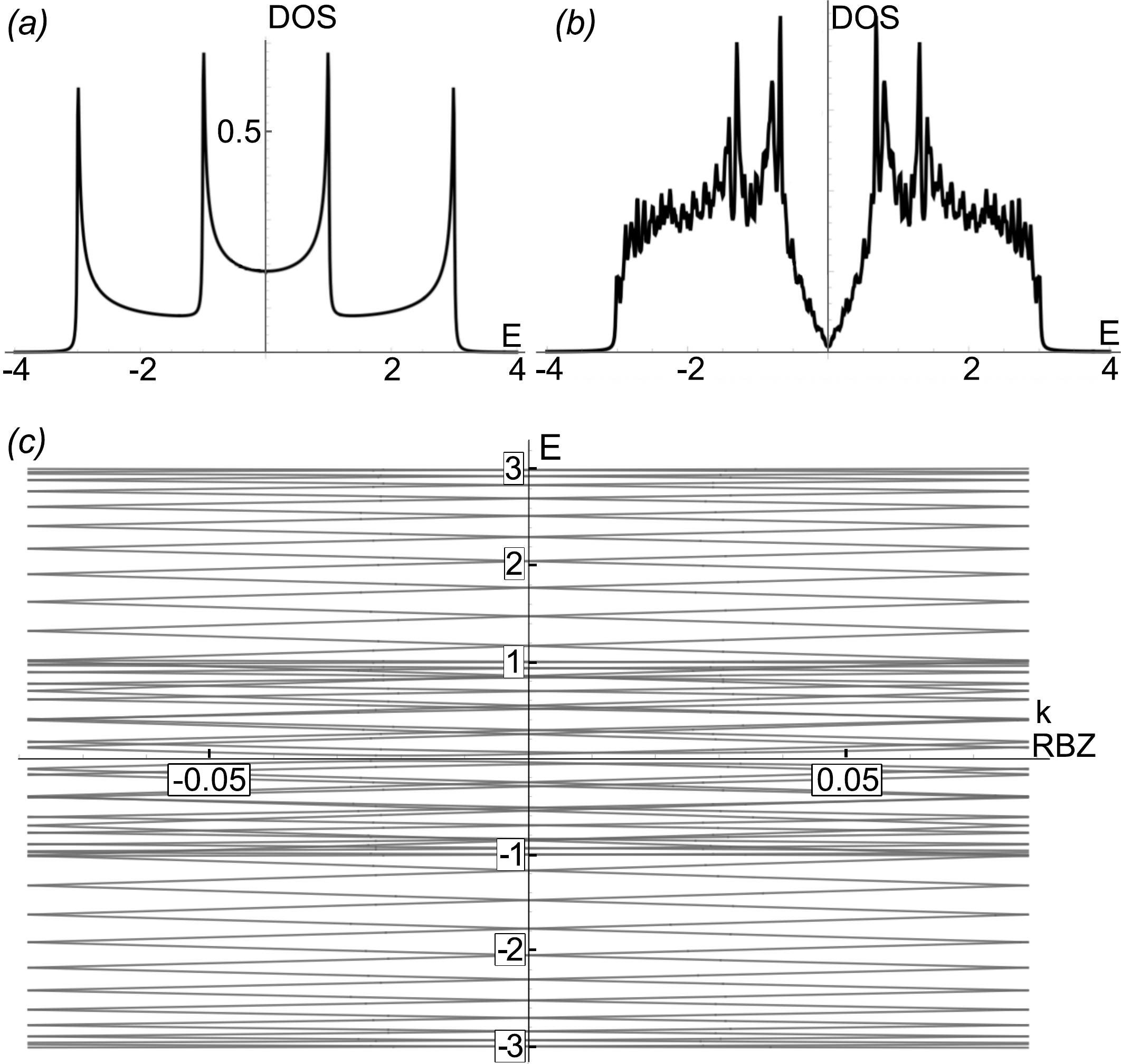}
\caption{\label{f2}DOS at T=0 and (a) $\delta=1$, (b) $\delta=\frac{17}{20}$. (c) Band spectra of the minibands $H_{\rm RBZ}$, in the RBZ.}
\end{figure}
The Fourier-transformed fermion operator of site $\alpha$ on leg $\ell$ is \(c_{\ell,\alpha,k}=\frac{1}{\sqrt{N}}\sum_n e^{-2ikn} c_{\ell,\alpha,n}\), where $n$ indexes composite unit cells and N is the total number of unit cells. Structure factors associated with leg-1 and leg-2 are respectively, \(c(k)=2\cos k\), \(c_\delta(k)=2\cos\!\big(k-\pi\delta\big)\), and the Peierls phase from the dimerization of leg-2 is $\phi(k)=k(1-\delta)+\pi\delta$. In Fourier space, the moiré rung modulation becomes \(t_\perp(k)=1+(1-\delta)\cos\!\big(qk\big)\). The kinetic Hamiltonian is $H=\sum_k \Psi_k^\dagger\,H(k)\,\Psi_k$, 
with 
$\Psi_k=(c_{1,A,k},c_{1,B,k},c_{2,A,k},c_{2,B,k})^\top$, and
\begin{equation}
\label{eq:H}
H(k)=
\begin{pmatrix}
0 & c(k) & t_\perp(k) & 0 \\
c(k) & 0 & 0 & t_\perp(k) \\
t_\perp(k) & 0 & 0 & c_\delta(k)\,e^{+i\phi(k)} \\
0 & t_\perp(k) & c_\delta(k)\,e^{-i\phi(k)} & 0
\end{pmatrix},
\end{equation}
The Hamiltonian matrix of Eq.\ref{eq:H} is given in the extended Brillouin zone (EBZ), \(\rm k\in (-qG_0,qG_0 )\),  where all $q$ momentum replicas generated by the BZ folding are unfolded and displayed explicitly. This leads to the momentum range $k\in(\rm -qG_0,qG_0)$ used in Fig.~1(b). The EBZ does not represent a physical Brillouin zone; rather, it is a convenient representation that makes the moiré-induced band folding and hybridization transparent. The physically relevant description is recovered by folding the EBZ into the reduced Brillouin zone of width $G_0/q$. 
For a given $\delta$, the similarity transformation $\Delta=\pi q$  applied to $H(k)$ imposes constraints on p and q that determine the generic global period $2\pi q$ that gives rise to EBZ.
On the other hand, the reduced first Brillouin zone of the moiré lattice (RBZ) spans the interval \((-b_s/2,b_s/2)\), and the reciprocal superlattice 
\(G_s = \{ m\,b_s\ |\ m\in\mathbb{Z}\}\). 
\begin{figure}
\centering
\includegraphics[width=\linewidth]{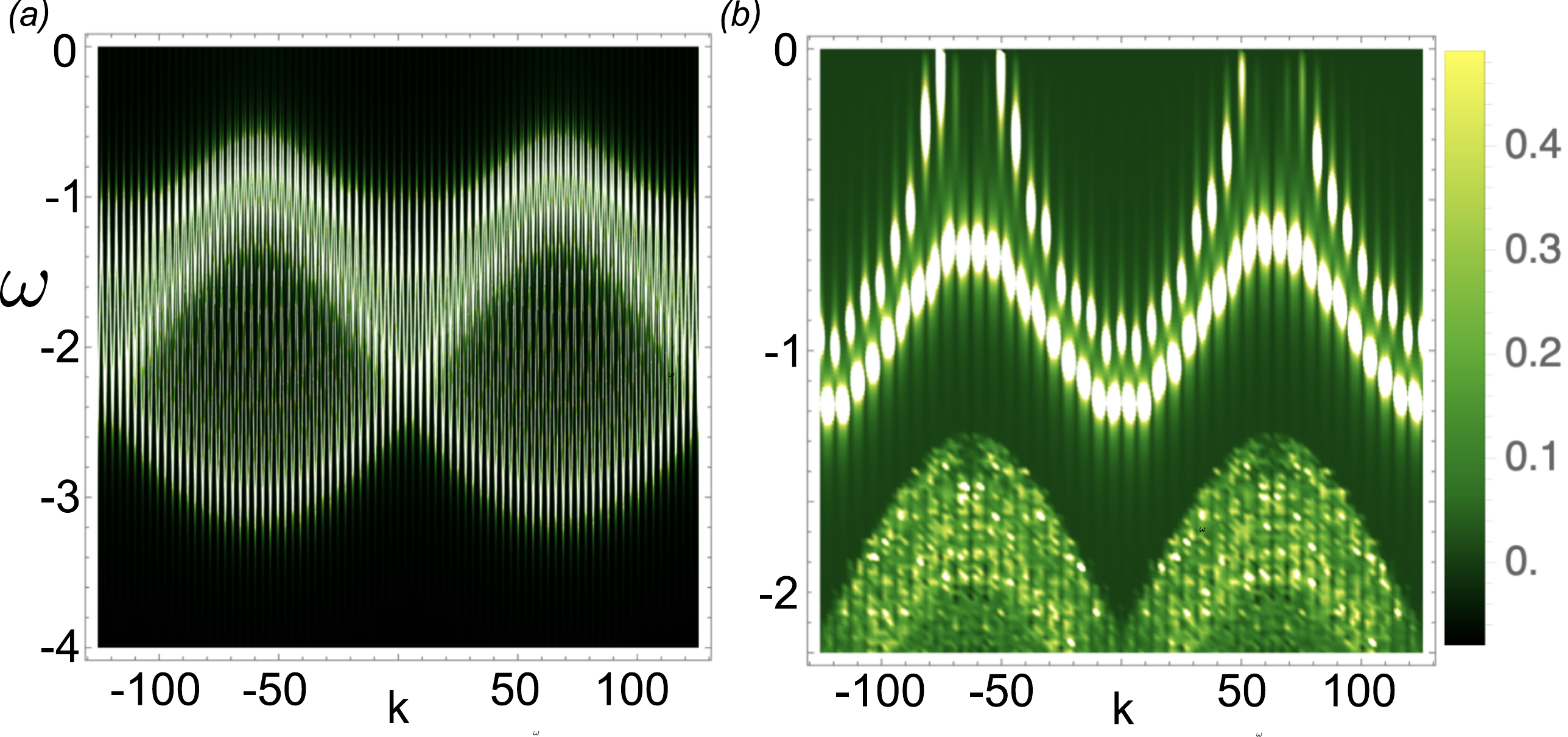}
\caption{\label{f3} Spectral function at $\delta=\frac{19}{20}$ and $T=1-\delta$ computed from (a) the single particle retarded Green function, (b) the two-particle RPA susceptibility.}
\end{figure}

To obtain moiré minibands in RBZ we keep the \(q\) replica momenta \(\{\bar{k}+m G_s\}_{m=0}^{q-1}\) in the Hamiltonian matrix $H_{\mathrm{RBZ}}$. Thus, the block‐diagonal piece becomes
\(
\big[H^{(0)}_{\mathrm{RBZ}}(\bar{k})\big]_{m m'}=
\delta_{m m'}\,H\!\big(\bar{k}+m G_s\big),\) 
\( m,m'=0,\dots,q-1
\). Rung modulation \(\cos(qk)\) connects nearest-neighbor replicas in the RBZ and gives rise to the off-diagonal part \(\big[\delta H_{\mathrm{RBZ}}(\bar{k})\big]_{m,m\pm 1}=\frac{t_1}{2}\,R,\quad
R = \sigma_1 \otimes \mathbb{I}_2\), where $\sigma_j$ is the j-$th$ Pauli matrix. The  \(4q\times4q\) moiré Hamiltonian in the RBZ becomes
\(H_{\mathrm{RBZ}}(\bar{k})=H^{(0)}_{\mathrm{RBZ}}(\bar{k})+\delta H_{\mathrm{RBZ}}(\bar{k})\).
If \(t_1=0\) the \(4q\) bands split into \(q\) independent copies of the four parent bands evaluated at $\bar{k}+m G_s$. The symmetry of \(H\) is preserved by $ H_{\mathrm{RBZ}}$, so a narrow band near \(E=0\) survives
for a broad window of parameters, as shown in Fig.\ref{f2}c.

The moiré model of Eq.\ref{eq:H} retains the two-leg/two-sublattice structure and captures the moiré-periodicity of the rung tunneling. The spectrum, exactly solvable, is shown in Fig.\ref{f1}(b-d). A tiny gap around the Fermi energy, E=0, is a direct consequence of \(\delta\), which breaks the translational symmetry.  Minigaps at folded-band crossings appear and scale linearly with $t_1$ at small modulation (large $\delta$). The position and width of these flat bands, as well as the position of the peaks in the density of states (DOS), can be tuned by $\delta$. With a perfect match between legs, the DOS depicts a pair of van Hove singularities (VHS) at the edges of the bands, a finite density of states at all energies, and another pair of VHS at $ E=\pm t_1$, signaling the coupling between the two legs, Fig.\ref{f2}(a). Once $\delta\neq 1$, the moiré potential produces three effects: 1) peaks at the edges of the spectra are smeared out; 2) the VHS at $E=\pm t$ splits into two peaks about $E\sim\pm t_1$  due to the breaking of inversion symmetry; and 3) the DOS $\sim \eta$ at the Fermi energy, as the system becomes gapped (Fig.\ref{f2}(b)). While Fig.\ref{f2}(b) appears symmetric in energy, the fine structure within the bands exhibits nontrivial modulation arising from the mismatch in periodicity.  $\delta$ and $t_\perp$ determine the overall shape and width of the energy bands and, consequently, the distance between the VHS. 

\emph{Interactions}
Throughout this work, we distinguish between the primary lattice-driven modulation, which generates the moiré structure discussed above, and a secondary electronic instability in the form of charge density modulations that may develop within the resulting miniband spectrum.
To study the nature of such charge density modulations \cite{coddens2006problem} and their associated collective excitations \cite{tucker1988theory} in our system, we included short-ranged Coulomb interactions on top of $H(k)$.  Following the standard approach, we constructed the low-energy Lagrangian using a path-integral method \cite{mahan2013many}. In this formulation, the trace over Hilbert space operators is re-expressed as an integral over a set of Grassmann variables \cite{mahan2013many}. The Hubbard-Stratonovich transformation (HS) \cite{altland2010condensed,tsvelik2007quantum} replaces the four-fermion interaction term with a term that couples the fermions to a fluctuating auxiliary bosonic field, making the path integral Gaussian and solvable.

We consider intra-leg Coulomb interactions between sites A and B within the same unit cell of equal strength $u$ in both legs. The interacting Hamiltonian is
\(
H_{\text{int}}
=\frac{u}{L}\sum_{\ell,\kappa}  \big[\,\rho_\ell(-\kappa)\,\rho_\ell(\kappa)\big],\)
with $\rho_\ell(\kappa)=\sum_{k}\sum_{s\in\ell} c^\dagger_{k+\kappa,s}\,c_{k,s}$.  At the saddle point, the order parameter $\Delta_k\equiv \Delta_0\,F(k)$ is time independent, and we choose $\Delta_0$ to be real. Restricting to the dominant bands $(m,n)$, the form factor is separable $
F(k)\equiv \sum_\ell\alpha_\ell\,\Gamma^\ell_{nm}( k,Q)$ where $\Gamma^\ell_{nm}$ denotes the leg-$\ell$ projected density vertex computed using the spectrum of $H(k)$. The mean-field Hamiltonian becomes
\(
\mathcal H_{\text{MF}}(k)=E_+(k)\sigma_0+E_-(k)\sigma_3+\Re\Delta_k\,\sigma_1-\Im\Delta_k\,\sigma_2,
\)
where $E_\pm(k)=\frac{E_m(k)\pm E_n(k{+}Q)}{2}$. The quasiparticle spectrum $\tilde{E}(k)=E_+(k)\pm\sqrt{|\Delta_k|^2+E_-^2(k)}$ consists of two branches with energy separation $2|\Delta_k|$  representing the CDW gap opened by electron-hole mixing. The single-particle retarded Green's function $G^R(k,\omega)$ describes the propagation of electron-like excitations with momentum $k$ and real frequency $\omega$ in an interacting system. The associated spectral function $ A(k,\omega)=
-\frac{1}{\pi}\,\Im\,G^{R}(k,\omega)$  measures the density of available states at momentum $k$ and energy $\omega$, and defines the ARPES intensity shown in Fig.\ref{f3}(a) for $\delta=19/20$.  The two bright arcs (upper and lower) correspond to the dispersions of the quasiparticles in our system, while the dark area between them is the gap region where no spectral weight exists. 

In the leg channel, the interaction matrix is $V=u\,\mathbb{I}_{2}$.
After expanding the term \textit{Trln} to quadratic order in the effective action, it becomes the Gaussian RPA action, whose functional derivative gives the interacting susceptibility $
\chi(\kappa,i\omega_n)$, where $\omega_n$ is the bosonic Matsubara frequency, $\omega_n=2\pi n T$. The CDW instability is determined from the static susceptibility
$\chi^{(0)}(\kappa,T)\equiv\chi^{(0)}(\kappa,i\omega_n=0;T)$. At a given temperature T, a density-wave occurs when 
$\det\!\big[\mathbb{I}_2-V\,\chi^{(0)}(\kappa,T)\big]=0
$. Letting  \(\lambda_{\max}^{(0)}(\kappa)\) be the largest eigenvalue of \(\chi^{(0)}(\kappa)\), the instability criterion reads $u\,\lambda_{\max}^{(0)}(\kappa,T)=1,\quad
Q*=\underset{\kappa}{\arg\max}\ \lambda_{\max}^{(0)}(\kappa,T)\ 
$. \(Q*\) is the instability wavevector at which the largest eigenvalue of the leg-projected bare matrix is maximal.  It sets the periodicity of charge modulation in reciprocal space. When $u\lambda>1$ we found that the system is unstable to CDWs at the wavevector $Q^*=p\pi$. Likewise, if at a given T, $u>u_c$, with the critical coupling $u_c=1/\lambda_{\max}^{(0)}(Q*,T)$, the system is unstable to the formation of a charge density wave phase. We found that at low temperatures, $T<t$, $u_c$ rises quadratically with $ T$, while at higher temperatures, it grows roughly linearly. 

The unstable mode sitting at $Q*$, which is not an integer multiple of a moiré reciprocal vector, indicates that, relative to the moiré lattice, this is an incommensurate CDW. In addition, projecting $\chi$ onto each leg at Q* we found that the RPA-unstable eigenmode exists equally on both legs, and that it is odd.  As the net charge modulation on a rung cancels to leading order, the phasons in our model are neutral \cite{wilson1985charge,fisher1985sliding,brazovskii2004pinning,PhysRevLett.42.1423}. 

\emph{Moiré Phason}
Collective modes of the CDW state manifest as poles of the two-particle RPA susceptibility. The retarded, leg-resolved Lindhard function is computed using leg projected vertices from the quasiparticle spectrum at $T<T_c$,
\begin{widetext}
\begin{equation}
\Pi^{R}_{XY}(\kappa,\omega)
= \frac{1}{N_k}\sum_{k}\sum_{n,m}
\frac{ f\!\big(\tilde{E}_m(k)\big) - f\!\big(\tilde{E}_n(k{+}\kappa)\big) }
     { \omega + i\eta + \tilde{E}_m(k) - \tilde{E}_n(k{+}\kappa) }\,
\tilde{\Gamma}^{X}_{nm}(k,\kappa)\,\tilde{\Gamma}^{Y}_{nm}(k,\kappa)^{*}
\label{eq:bubble}
\end{equation}
\end{widetext}
where $f$ denotes the Fermi-Dirac distribution and $\eta$ is a small broadening parameter. The retarded RPA susceptibility $
\chi^{R}(\kappa,\omega)=\Big[\mathbb{I}_2 - V\,\Pi^{R}(\kappa,\omega)\Big]^{-1}\,\Pi^{R}(\kappa,\omega)$
gives rise to the two-particle spectral function of Fig.\ref{f3}(b) which depicts three regimes. The lowest domes $(\omega\lesssim -1.3)$ correspond to the particle–hole continuum due to the imaginary part of the retarded bubble, \(\Im \Pi^{R}(\kappa,\omega)\), which gives the density of states of pair-hole pair excitations. It measures the energy necessary to create a particle–hole pair by removing an electron from band n with momentum k and adding an electron to band m with momentum $k+Q^*$. The middle bright dispersive band \((-1.2\lesssim\omega \lesssim -0.8)\) corresponds to the amplitudon, a Higgs-like oscillation of the magnitude of the excitonic order parameter. The upper bright, sharp dispersive branch \((-0.7\lesssim \omega \lesssim 0)\) shows the poles of the RPA susceptibility that produce the gapless phason. The CDW ordering vector is \(Q^{*}\), so the Goldstone mode softens at  \(\kappa\sim Q^{*}\).

To examine the phason dynamics,  we promote the order parameter to a slowly varying
complex field, \(\Delta_0 F(k)\,e^{i\theta(\tau,x)}\), with a slowly varying phase and fixed amplitude $\Delta_0$. The unitary rotation \(R_\theta=e^{+\frac{i}{2}\theta\tau_3}\), removes the phase
from the gap term, and, to linear order in gradients, couples minimally the fictitious gauge field \(a_\mu=\tfrac12\partial_\mu\theta\) to the
fermions. 
Integrating out the gapped fermions and expanding the fermionic determinant to second order in $a_\mu$ gives the action for the phason in the long-wavelength, low-frequency limit.
\begin{multline}
S_{\rm ph}=\frac12 \int d\tau dx\;
\Bigl[
K\,(\partial_\tau\theta)^2+\rho\,(\partial_x\theta)^2
\Bigr],\\\omega^2(k)=\nu_{\theta}^2 k^2,\;\;
\nu_{\theta}^2=\rho/K.
\end{multline}
with \(K\) and \(\rho\) the dynamic and static stiffness of the phason, respectively, and $\nu_\theta$ the phason speed. For weak coupling, at $T<T_c$ and at $k\sim Q*$, a gradient expansion of the effective action in time and space retains terms $\omega^2$ and $k^2 $ with BCS coefficients \cite{chaikin1995principles,gruner1988dynamics}
$
K=\frac14\sum_{k,b}\frac{|\Delta_{k,b}|^2}{E_{k,b}^3}\tanh\frac{E_{k,b}}{2T}
$, and $
\rho=\frac{1}{2}\sum_{k,b}v_{-}^{2}\frac{|\Delta_{k,b}|^2}{E_{k,b}^3}\tanh\frac{E_{k,b}}{2T}
$, where $v_{-}=\partial_k E_{-}(k,b)$, $|\Delta_{k,b}|^2=|\Delta_0 F_-(k)|^2$, and $F_-(k)$ are the form factors projected in the odd channel.
$\Delta_0$ can be estimated from the form factors evaluated at the \textit{hot-spot} momentum $k*$, which marks the points of the Fermi surface connected by $Q*$ with the largest susceptibility. $k*$ is found by minimizing $|E_m(k)-E_n(k+Q*)|$ with respect to k and identifying the relevant bands. At $\delta=\frac{19}{20}$, we found $m=n=4$, $k*\sim 21.3$, and $\Delta_0\sim 0.7$ with the intraband instability arising from the inclusion of interactions that preserve band independence. Having $\Delta_0$ we find $\rho\sim 0.32$ (in units of $t$), and  $K\sim 6.7$ (in units of $t^{-1}$). As $\rho/K\sim 0.05$; this is a soft mode. The phason speed $\nu_\theta\sim 0.2-0.17 i$, in units of $\frac{a t}{\hbar}=v_F$, where $v_F$ is the Fermi velocity. For typical energy and length scales $t\sim 0.1-1$ [eV], $a\sim 3-5$ [A], we obtain $\nu_\theta\sim 2\times 10^4-2\times 10^5$ [m/s]. The damping ratio $\operatorname{Im}[\nu_\theta]/\operatorname{Re}[\nu_\theta]\sim 0.05$; therefore, for the slight inter-leg mismatch, these soft phasons are long-lived.  

Near the hot spots, one can estimate how \(\nu_\theta\) scales with t and $\delta$. In our model, minibands are a consequence of the moiré rung-modulation, so the effective moiré potential scales like $V^{\mathrm{eff}}_{\mathrm{moire}} \sim (1-\delta)t$. In perturbation theory, the miniband width scales like 
\(W(\delta)\sim \frac{V_{\text{moire}}^{\text{eff}\,2}}{\Delta E}
\) and in the weak-coupling limit, $W(\delta)\sim V_{\text{moire}}^{\text{eff}}$.  In the CDW phase, \(\rho\) is roughly determined by the cost of twisting the phase in space, which is controlled by the effective DOS, \(N_{\mathrm{eff}}\), and the relevant Fermi velocity of the low-energy band that condenses, \(\rho\sim N_{\mathrm{eff}} v_F^2 \). The relevant Fermi velocity in the miniband comes from the miniband width
$v_F \sim W(\delta) \propto (1 - \delta)\, t$. 
As minibands flatten when \(\delta\to 1\),  \(N_{\mathrm{eff}}\sim const/t\). That gives \(\rho \sim (1 - \delta)^2 t\).
Phason inertia \(K\) arises from the $\omega$-dependence of $\Pi^R(k,\omega)$ at \(k\to 0\), and the scaling is dominated by \(
K\sim 1/t\). Therefore, \(
\nu_{\theta}(\delta,t)\sim t(1-\delta)\). Increasing $t$ raises \(v_F\) and with that,
\(\rho\) and \(\nu_{\theta}\). On the other hand, \(K\) decreases with $t$
as \(1/v_F\).
Altogether, we conclude that increasing $t$ speeds up phasons, while pushing $\delta\to 1$ slows them down.

\begin{figure}
    \centering
    \includegraphics[width=\linewidth]{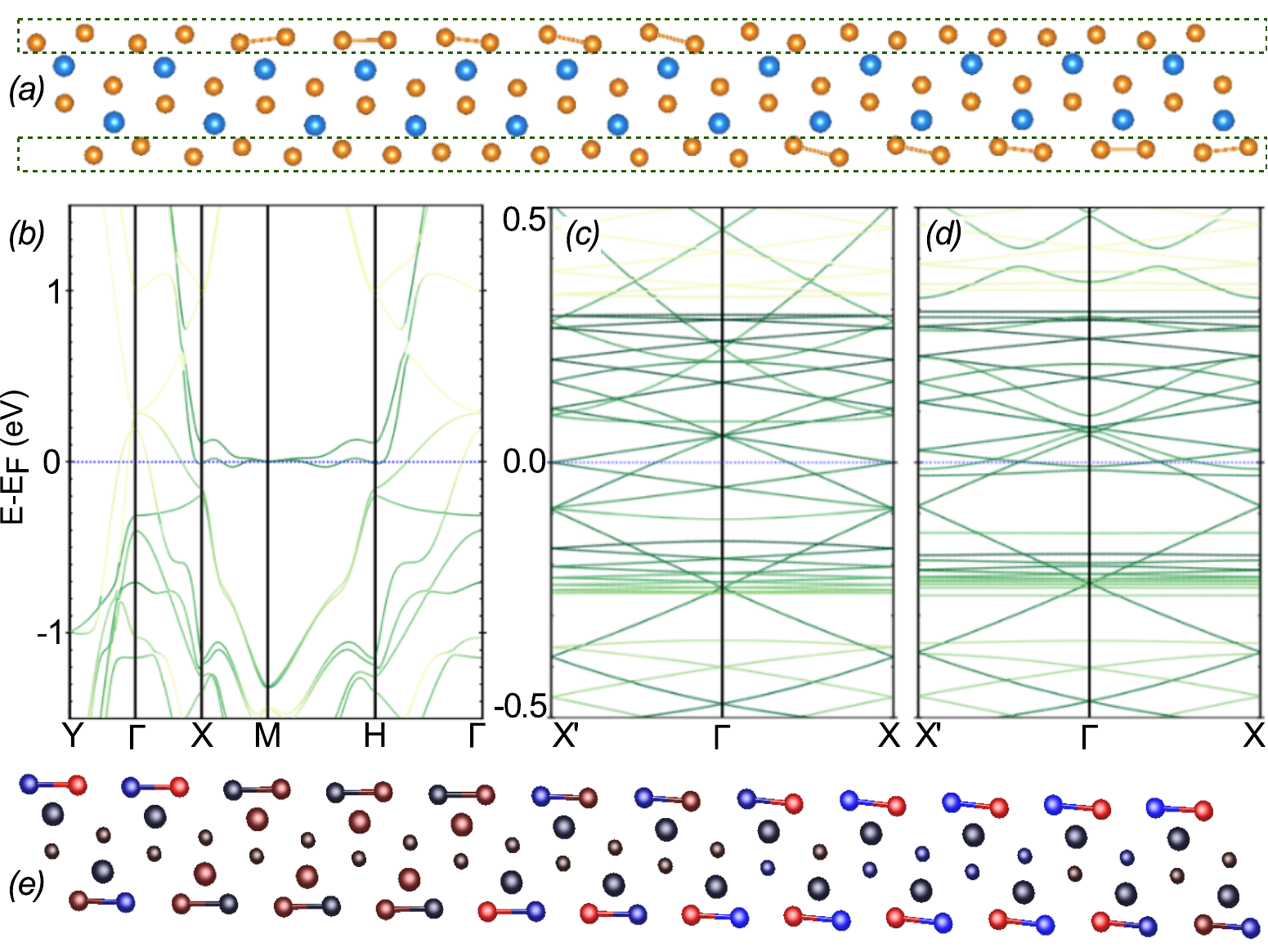}
    \caption{(a) Distorted structure of HfTe$_3$, the distortion is exaggerated, Te chains are highlighted. (b) Bulk bandstructure; the darker the color, the stronger the projection on the Te chains. Subbands of a $12\times1$ supercell of a HfTe$_3$ monolayer, without (c) and with (d) the modulation. (e) Changes of the charge around each atom, bright blue/red is $\pm0.002e$.}
    \label{fig:dft_short}
\end{figure}

\emph{Hafnium Tritelluride} The moiré-based framework developed here is naturally suited to quasi-1D materials with chain-like structural motifs. As a concrete example, we examine the transition-metal trichalcogenide $\mathrm{HfTe_3}$, whose bulk displays charge-density-wave order \cite{denholme2017coexistence, liu2021quasi}. DFPT calculations on the surface reveal a phonon instability at  $q\simeq0.06\mathbf{b}^{\ast}$ involving out-of-plane motion of the dimerized Te chains, generating a long-wavelength structural modulation (Fig.~\ref{fig:dft_short}a). This distortion is accommodated in a $12\times1\times1$ supercell, whose relaxed geometry develops a chain-dependent phase shift—an explicit real-space moiré pattern emerging at negligible energetic cost. A Wannier analysis shows dominant intra-chain dimerization and weak inter-chain tunneling, motivating a reduced strong-coupling model with chain-dependent oscillatory hoppings. The resulting moiré Hamiltonian reproduces the flat minibands, band reconstruction, and gap openings obtained in the DFT supercell (Fig.~\ref{fig:dft_short}c-d), demonstrating that the instability-driven modulation reshapes the low-energy electronic structure in direct analogy with moiré systems. The changes in the charges due to modulation (Fig.~\ref{fig:dft_short}e) have a dipolar pattern consistent with long-range neutral excitations.These results highlight chain-dependent phase shifts and structural incongruities as a microscopic pathway to electronic instabilities in quasi-1D materials.

\emph{Conclusions}
In this work, we demonstrate that the ordered phase of the moiré-modulated two-leg ladder realizes an excitonic charge-density wave. Microscopically, it is characterized by a particle–hole condensate formed between two flat minibands. Rung modulation produces a small indirect band gap, which permits interband particle–hole pairing, analogous to the mechanism of excitonic insulators in multiband semimetals. At the mean-field level, the resulting Hamiltonian is algebraically equivalent to a BCS Hamiltonian, except that the usual pairing sector is replaced by particle–hole coherence. Because the CDW in this electronic model does not couple to the lattice and breaks a continuous symmetry, the system hosts a neutral acoustic phason, which is the Goldstone mode of the excitonic condensate.  The dispersions of the CDW collective modes, the phason and the amplitudon, appear in the two-particle spectral function computed from the retarded susceptibility, whose imaginary part also depicts the CDW continuum at higher energies.

As the effective interleg modulation amplitude is proportional to $(1-\delta)$, both the miniband width and the relevant Fermi velocity scale as $(1-\delta)t$, the phason stiffness scales as $(1-\delta)^2 t$, whereas the phason inertia maintains the $1/t$ scaling, up to weak $\delta$-dependent corrections. Consequently, the phason velocity scales as $t(1-\delta)$.

In closing, we note that when leg-2 becomes nearly fully dimerized, the minibands disappear because the low-energy spectrum develops an almost perfectly flat bonding–antibonding structure,
\(E_{\pm}(k) \sim \pm t \delta + \mathcal{O}(1 - \delta)\). One of the bands thus becomes essentially dispersionless, and the rung-mediated hybridization with leg-1 is strongly suppressed: the cosine modulation is then too weak to generate the momentum-space replicas required to narrow the minibands. The intra-cell splitting of the dimerized chain is \(\Delta_{\mathrm{dimer}} \sim 2 t \delta\), and once this gap is large, the modulation can no longer efficiently mix the states.

The DFT and DFPT calculations presented here reveal an intrinsic structural instability associated with a soft phonon mode, which generates a long-wavelength, chain-resolved modulation of the electronic hopping amplitudes. This modulation generates a moiré-like electronic structure that forms the foundation of our effective model. Therefore, the excitonic CDW examined here should be interpreted as a secondary, interaction-induced electronic instability that can emerge once the moiré minibands have formed and the electronic kinetic energy has been substantially suppressed. At present, there is no direct experimental or first-principles evidence that such an excitonic CDW is realized in HfTe$_3$. Rather, our results demonstrate that the moiré modulation generated by a lattice instability provides a natural and robust platform for stabilizing an incommensurate electronic CDW with a neutral phason mode. In this sense, HfTe$_3$ serves as a physically motivated example showing how lattice-driven moiré patterns can enable emergent electronic collective phases at lower energies.
\begin{acknowledgments}
P.M. acknowledges support from the Fondo Nacional de Desarrollo Científico y Tecnológico (Fondecyt) under Grant No. 1250122 and from the Indian Institute of Technology Madras, India, where part of this work was conducted. J.C.E acknowledges ANID for her national doctoral scholarship No. 21231429 and partial support from UAI. F.M.  acknowledges support from Fondecyt grants 1231487 and 1220715, CEDENNA CIA250002, and partial support by the supercomputing infrastructure of the NLHPC (CCSS210001). This work used Bridges-2 at Pittsburgh Supercomputing Center through allocation PHY150003P from the ACCESS program.
\end{acknowledgments}
%\bibliography{paper_phason}% Produces the bibliography via BibTeX.

\end{document}